\documentclass[10pt,sigconf]{acmart}

\usepackage[english]{babel}

\usepackage{enumitem}
\usepackage{cleveref}
\usepackage{tabularx}
\usepackage{ragged2e}  %
\newcolumntype{Y}{>{\RaggedRight\arraybackslash}X}

\renewcommand\footnotetextcopyrightpermission[1]{} %
\setcopyright{none}
\acmDOI{}

\acmISBN{}

\acmPrice{}

\usepackage{subfigure}
\usepackage{graphicx}
\usepackage[export]{adjustbox}
\usepackage{amsmath}
\usepackage{multicol,multirow}
\usepackage{epsf,psfrag,pstricks}
\usepackage{mdwlist}
\usepackage{rotating}
\usepackage{xspace}

\usepackage{url}
\usepackage{breakurl}

\usepackage{capt-of}
\usepackage{flushend}

\newcommand{\ie}{i.e., \@}
\newcommand{\eg}{e.g., \@}

\newcommand{\etal}{et al.\xspace}

\usepackage{fancyhdr}
\fancypagestyle{plain}{%
   \fancyhf{} %
    \fancyfoot[L]{\small{}}%
    \fancyfoot[R]{\small{}}%
}
\def\refname{}
\fancypagestyle{firstpagestyle}{%
   \fancyhf{} %
   \fancyfoot[L]{\small{}}%
   \fancyfoot[R]{\small{}}%
   \fancyfoot[C]{\thepage}
}

\usepackage[absolute,showboxes]{textpos}

\begin{document}

\pagestyle{fancy}   %

\renewcommand{\sectionautorefname}{\S}
\renewcommand{\subsectionautorefname}{\S}

\title{One Bad Apple Can Spoil Your IPv6 Privacy} %

\author{Said Jawad Saidi}
\affiliation{%
  \institution{MPI-INF/Saarland University}
}
\email{jsaidi@mpi-inf.mpg.de}

\author{Oliver Gasser}
\affiliation{%
  \institution{MPI-INF}
}
\email{oliver.gasser@mpi-inf.mpg.de}

\author{Georgios Smaragdakis}
\affiliation{%
  \institution{TU Delft}
}
\email{g.smaragdakis@tudelft.nl}

\renewcommand{\shortauthors}{} %

\renewcommand{\shortauthors}{} %

\pagestyle{plain}

\begin{abstract}

IPv6 is being more and more adopted, in part to facilitate the millions of smart devices that have
already been installed at home. Unfortunately, we find that the privacy of a
substantial fraction of end-users is still at risk, despite the efforts by
ISPs and electronic vendors to improve
end-user security, e.g., by adopting prefix rotation and IPv6 privacy extensions.
By analyzing passive data from a large ISP, we find that around
19\% of end-users' privacy can be at risk. When we investigate the root
causes, we notice that a single device at home that encodes its MAC address into the IPv6 address can be utilized as a tracking identifier for the
entire end-user prefix---even if other devices use IPv6 privacy
extensions. Our results show that IoT devices contribute the most to this privacy
leakage and, to a lesser extent, personal computers and mobile devices. To our
surprise, some of the most popular IoT manufacturers have not yet adopted
privacy extensions that could otherwise mitigate this privacy risk.
Finally, we show that third-party providers, e.g., hypergiants, can track up to
17\% of subscriber lines in our study.%

\end{abstract}

\maketitle

\setlength{\TPHorizModule}{\paperwidth}
\setlength{\TPVertModule}{\paperheight}
\TPMargin{5pt}
\begin{textblock}{0.8}(0.1,0.02)
    \noindent
    \footnotesize
    If you cite this paper, please use the SIGCOMM CCR reference:
    Said Jawad Saidi, Oliver Gasser, and Georgios Smaragdakis. 2022.
    One Bad Apple Can Spoil Your IPv6 Privacy.
    In \textit{ACM SIGCOMM Computer Communication Review, Volume 52, Issue 2, April 2022.}
    ACM, New York, NY, USA, 9 pages.
\end{textblock}

\section{Introduction}\label{sec:intro} %

The adoption of IPv6 in the Internet is continuously increasing~\cite{IPv6-state-2021}. One of the drivers is the
unprecedented demand for smart devices at home, ranging from voice assistants
to smart TVs and surveillance cameras, that all have to be assigned addresses to
have access to the Internet and the cloud~\cite{APNIC-IPv6-IoTs}. While the use of Network
Address Translation (NAT) and concerns about %
IPv6 addressing privacy have delayed its adoption, operators, vendors, and
the research community have long ago provided privacy solutions to mitigate these
risks. ISPs have adopted prefix rotation~\cite{rfc8415} and
network equipment manufacturers and software developers have enabled IPv6 privacy
extensions~\cite{rfc4941,rfc8981}. 

A recent work~\cite{rye2021follow} shows that if the home network gateway router, also referred
to as customer premises equipment (CPE), is using a legacy IPv6 addressing
standard employing EUI-64 (Extended Unique Identifier), it is possible 
to track devices that use IPv6 at home using active measurements. Unfortunately,
in this paper, we report that even if the CPE and the ISP apply best common
practices, \ie IPv6 privacy extensions and prefix rotation, it is still
possible to track devices that use IPv6 at home. In detail, we show that the
existence of only a single device that uses EUI-64 at home can spoil the privacy of
potentially all IPv6-enabled devices and eventually end users' privacy across
these devices. To estimate the risk in a realistic setting, we rely on passive
measurements, namely network flows collected at a large European ISP. However, any
third-party provider, such as hypergiants~\cite{SIGCOMM2021-hypergiants}, network
traffic aggregators (Internet exchange point, upstream providers), or service
providers (e.g., NTP, DNS providers), receiving connections from devices at
the same home can potentially defeat the privacy of current IPv6 solutions even
if only one these devices uses the legacy EUI-64 technique. Unfortunately, the
average end-user is not in a position to know which of their devices use EUI-64.

\noindent Our contributions can be summarized as follows:

\vspace{-.5em}
\begin{itemize}[leftmargin=*]

\item We perform a study at a large European ISP. Our analysis shows that around 19\% of
end-user prefixes host at least one device that does not use IPv6
privacy extensions.%

\item We show that the existence of even a single device without privacy extensions
in an end-user prefix can defeat the ISP-deployed prefix rotation and IPv6
privacy extensions adopted by hardware vendors to preserve user privacy.

\item Our analysis shows that the majority of devices without privacy extensions,
responsible for spoiling users' privacy, are devices of IoT manufacturers.
However, computer and mobile manufacturers are also contributing.

\item We show that, in most cases, a single device without privacy
extensions is responsible for privacy leakage. Unfortunately, these devices have
been manufactured by market leaders. Thus, it would have been possible to
prevent this privacy leakage if these manufacturers had adopted best common
practices, i.e., IPv6 privacy extension.   

\item We also show that a popular content provider, application, or service contacted by a
device that is not using privacy extensions can track the user and other
contacting devices across rotating prefixes. Unfortunately, the privacy of
up to 17\% of subscriber lines can face this risk.

\end{itemize}

\vspace{-.5em}

\section{Background}\label{sec:background} %

To solve the address shortage in IPv4 among other things, the networking
community introduced the IPv6 protocol more than two decades ago
\cite{rfc2460,rfc8200}.  Nevertheless, IPv6 is only recently being
deployed on a larger scale~\cite{karpilovsky2009quantifying} with about 36\% of
all requests to Google going over IPv6 as of March 2022 \cite{google-ipv6}.
In addition to the IPv6 address space being larger, the addressing itself is
also different compared to IPv4~\cite{rfc7707}.  While in IPv4 most end-user
clients get their address via DHCP~\cite{rfc2131}, in IPv6 clients get addresses
either via DHCPv6 \cite{rfc8415} or stateless address auto-configuration (SLAAC)
\cite{rfc4862}. Instead of directly assigning a full address as in DHCP or
DHCPv6, with SLAAC a router simply sends a prefix to its clients (\ie the
network part), and the clients then by themselves choose an IPv6 address within
that prefix (\ie the host part). This host part is also called interface
identifier or {\it IID}.  Initially, the IID part used an encoding of the interface's
MAC address, called \emph{EUI-64} \cite{ieee2018guidelines}.  The unique and
consistent nature of MAC addresses lead to devices being trackable over time
and across different networks \cite{rye2019eui}.  Consequently, IPv6
\emph{privacy extensions} were proposed, which simply randomize the IID part
instead of using a device's MAC address \cite{rfc4941}.  In addition to user
devices being trackable by EUI-64 addresses, ISP subscribers can also be
tracked by their prefix.  In order to defeat prefix tracking, ISPs can change
the prefix of each customer after a certain time (\emph{prefix rotation}).
Although there has been a lot of work on IPv6 measurements
\cite{gasser2016scanning,gasser2018clusters,rye2020discovering,rye2019eui,beverly2018ip,rohrer2016empirical,strowes2016,fiebig2017something,fiebig2018rdns,borgolte2018enumerating,ullrich2015reconnaissance,foremski2016entropy,murdock2017target,fukuda2018knocks,liu20196tree,bajpai2019longitudinal,almeida2020classification,li2020towards,padmanabhan2020dynamips,hou20216hit,cui20216gan,li2021fast,zheng2020effective,bruns2020network},
many of them focused on active measurements or structural properties of the
IPv6 space.  The work closest to ours was recently published by Rye \etal
\cite{rye2021follow}, in which they show that prefix rotation can be defeated by
tracerouting customer premise equipment (CPE), which responds with EUI-64
addresses.  In our work, %
we show the privacy
implications of EUI-64 usage among devices directly within the end-user network.

\section{Methodology} \label{sec:methodology} 

\begin{figure} [!bpt]
	\captionsetup{skip=.25em,font=small}
	\includegraphics[width=.95\linewidth,valign=t]{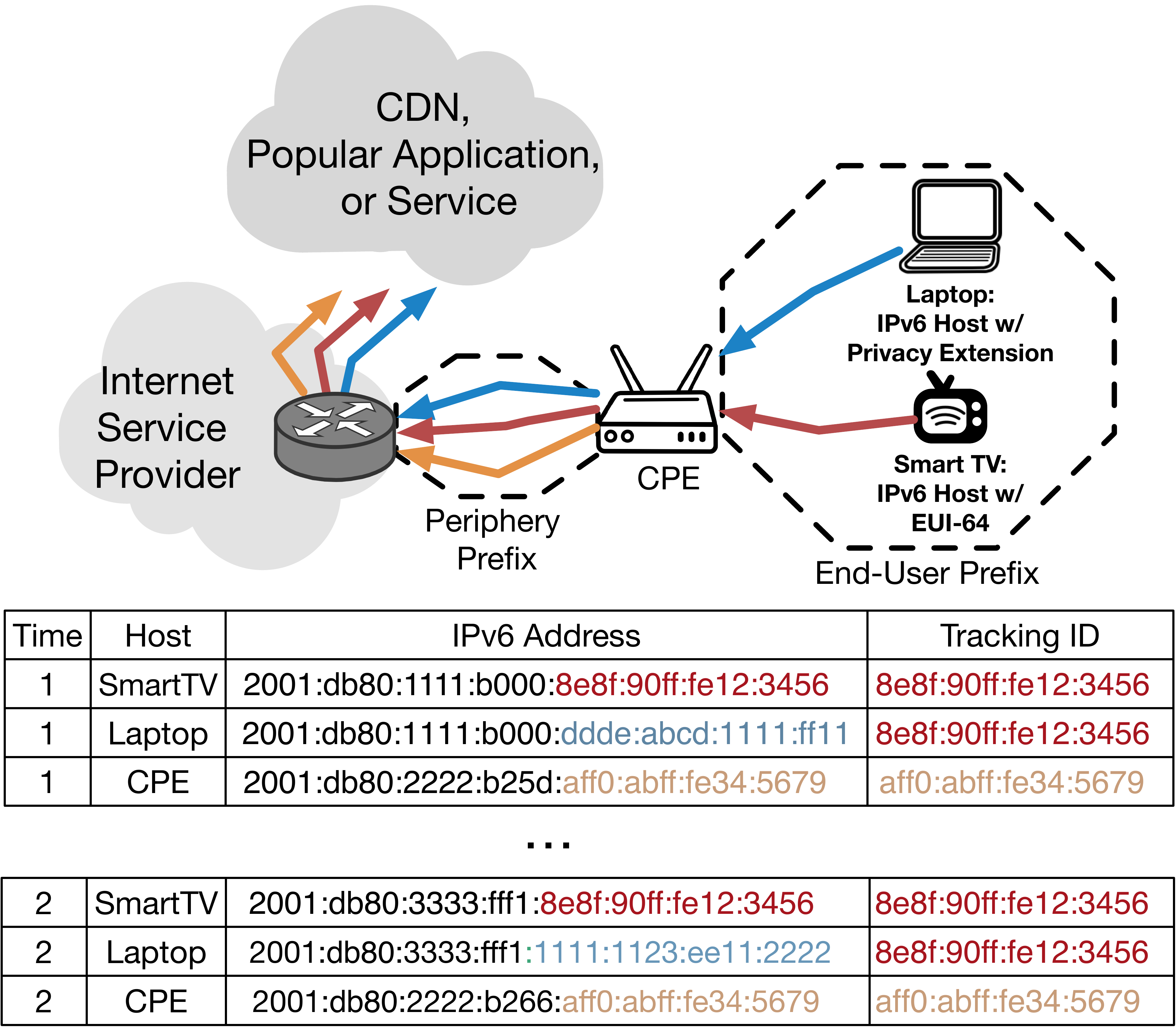}
	\caption{Privacy leakage across prefixes.}
	\label{fig:iot-privacy-leakage}
\vspace{-1em}
\end{figure}

In this section, we describe our methodology and show how a single device using
EUI-64, \ie not using privacy extensions, can be used to track devices at the
subscriber level.  In \Cref{fig:iot-privacy-leakage}, we show how an end-user
prefix can be tracked despite the ISP performing frequent prefix rotation.  In
the example scenario, there are two devices in the end-user prefix, a laptop and
a smart TV.  Both are using IPv6, the former with privacy extensions, the latter
with EUI-64.  The CPE device also has IPv6 connectivity on the upstream facing
interface.  If the CPE device's WAN-facing address is not within the end-user
prefix, it can not be used for tracking with our methodology.

Since the smart TV is not using privacy extensions it allows CDNs and other large players in the Internet to track not only the smart TV itself, but all devices within that end-user prefix.
In fact, we can use the smart TV's IID part of the IPv6 address as its unique
tracking ID since it is derived from a MAC address.  Furthermore, we assign this
same {\it tracking ID} to all addresses within the end-user prefix.  This way,
we can jointly track all devices of a subscriber by relying on a single
EUI-64-enabled device.
After the initial blue and red flows were observed, the ISP rotates the
customer's prefix (time 2), and all customer devices are now using a new IPv6
address.  Importantly, as the smart TV is still using the same IID even in this
new prefix, any content provider can again associate all devices with the same
tracking ID as before.  With this technique, a single EUI-64 device in an
end-user subnet can spoil the privacy gains of prefix rotation of all other
devices, even if they use privacy extensions.

For our method to be effective, the devices in the same end-user prefix must
contact a vantage point. %
In our case, we are in a privileged position to see all the connections and thus be able to track all the
devices. However, in the wild, these devices would require to contact the same
application, \eg hypergiants, content delivery networks, search engines,
upstream providers, or other popular services such as DNS or NTP.  The devices
can then simply be tracked by assigning tracking IDs to the red and blue flows
as shown in \Cref{fig:iot-privacy-leakage}.

Recall, the IID part of an EUI-64 IPv6 address is generated by inserting the {\tt
`ff:fe'} hex string between the third and fourth bytes of a MAC address and
setting the Universal/Local bit. We can extract the MAC address from the
EUI-64 part of an IPv6 address and uncover the device manufacturer.
To achieve this, we extract the Organization Unique Identifier (OUI) part of the
MAC address, \ie the first three bytes.  For the mapping, we use the official
IEEE OUI database~\cite{mac-oui}. This database contains information about the %
name and address of the manufacturer that has registered the OUI.

\section{Datasets}\label{sec:datasets} %

\noindent{\bf ISP Profile:} We analyze data from a large European
Internet Service Provider (ISP) that offers Internet connectivity to
more than 15 million broadband subscriber lines in Europe.

\noindent{\bf IPv6 Assignment at the ISP:} The ISP fully supports IPv6
by utilizing dual-stack addressing.  Each CPE device gets delegated a
/56 IPv6 prefix, out of which it will pick one /64 prefix, which is
then used to assign addresses to clients via SLAAC.
By default, the ISP rotates the /56 prefixes delegated to customers
every 24 hours.  Generally, the IPv6 prefix used for the upstream-facing CPE
interface to the ISP (``periphery prefix'' in
\Cref{fig:iot-privacy-leakage}) 
may or may not share the same prefix as the end-user network. Thus, in the latter case, a /56 prefix that does not contain an upstream-facing CPE interface represents an end-user network. We will show in our analysis in ~\Cref{sub:quantify} that the CPE interface and end-user networks of this ISP do not share the same /56 prefixes.       

\noindent{\bf ISP Data:} The data is sampled network flow data collected at the ISP
using NetFlow~\cite{Cisco-Netflow} to assess the state and operation of
its network routinely, a typical operation of ISPs. For our analysis, we apply our method
on the NetFlow data at the premises of the ISP, and we do not transfer or have
direct access to the NetFlow data. The data was collected on July
14, 2021, and four months later, on November 17, 2021.  

\noindent{\bf Ethical Considerations:} The ISP NetFlow data does not contain any
payload. Thus, there is no user information. The data is processed
on-premise at the ISP, and no data is copied, transferred, or stored outside the
server dedicated for NetFlow analysis at the ISP. Because IPv6 can be used as
Personal Identifiable Information (PII), we consistently hash the first 56 bits
that the subscribers of the ISP use. Following best operational practices, the
NetFlow data is deleted at an expiration date set at the data collection time.
To avoid blocklisting of products, vendors, manufacturers, and network companies,
including hypergiants, we anonymize the names of all companies.

\section{Privacy Violations at the Edge}\label{sec:detection} %

To assess the prevalence of privacy violations due to devices without
privacy extensions, we apply our methodology on NetFlow data of the
ISP (see the previous section). Since the ISP rotates the customer
prefixes once a day, we analyze one day of data, namely, Wednesday,
July 14, 2021, to show the feasibility of tracking devices at home. We
also examine the data collected on Wednesday, November 17, 2021, which
confirms our initial observations. Unless otherwise mentioned, our
results refer to the first dataset.

\subsection{Quantifying EUI-64 Prevalence}

\begin{figure} [t]
	\captionsetup{skip=.25em,font=small}
	\includegraphics[width=1\linewidth,valign=t]{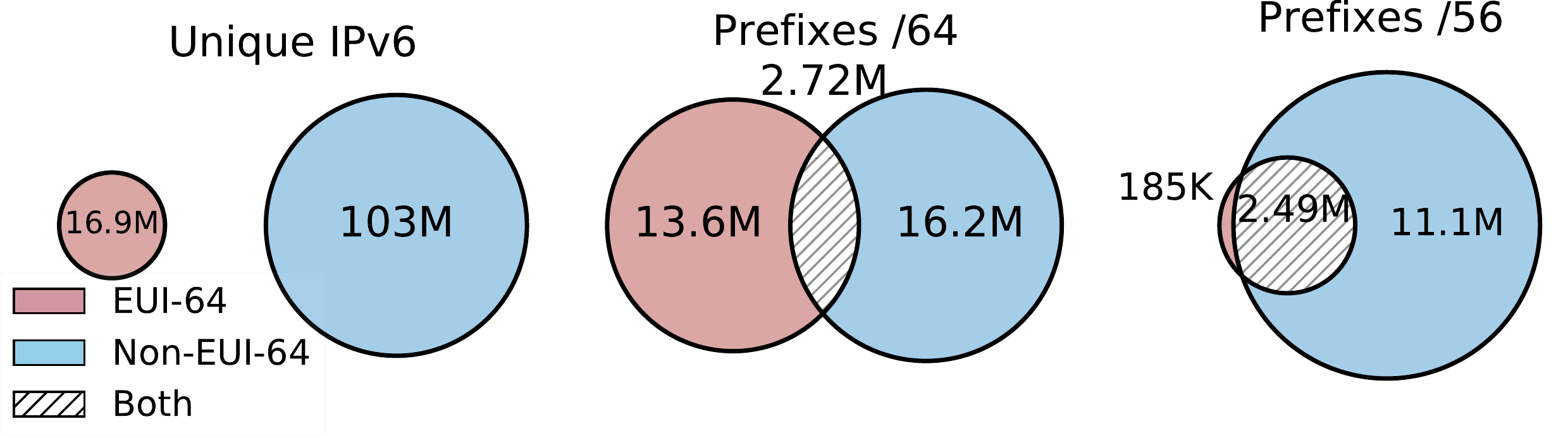}
	\caption{Venn diagram for EUI-64 and non-EUI-64 IPv6 addresses and the overlap between different prefix sizes.}
	\label{fig:isp-all-EUI64}
        \vspace{-0.5cm}
\end{figure}

In Figure~\ref{fig:isp-all-EUI64} (left), we report the number of IPv6
addresses visible in the ISP during one day.  Recall that the ISP
serves around 15 million subscriber lines. The number of non-EUI-64
addresses---in our case those are IPv6 addresses with privacy
extensions enabled (see \Cref{sec:hamming} for a detailed analysis of
non-EUI-64 addresses)---is more than 100 million.  This is to be
expected as these devices frequently use new IPv6 addresses, and more
than one of these devices may be served by a subscriber line. On the
other hand, the number of IPv6 addresses for devices that do not use
privacy extensions, i.e., EUI-64, is smaller, around 17 million.
However, we have strong and consistent identifiers for IPv6 addresses
used by these devices, i.e., their IIDs, that we use to track devices
even when the ISP performs prefix rotation. In total, we found 14.4
million devices that use EUI-64.

Next, we map all IPv6 addresses to their /64 prefix.  We see that the
numbers are now quite similar, 13.6M for EUI-64 and 16.2M for prefixes
with non-EUI-64 addresses, with an overlap of 2.7M prefixes.

\begin{figure} [t]
	\captionsetup{skip=.25em,font=small}
	\includegraphics[width=1\linewidth,valign=t]{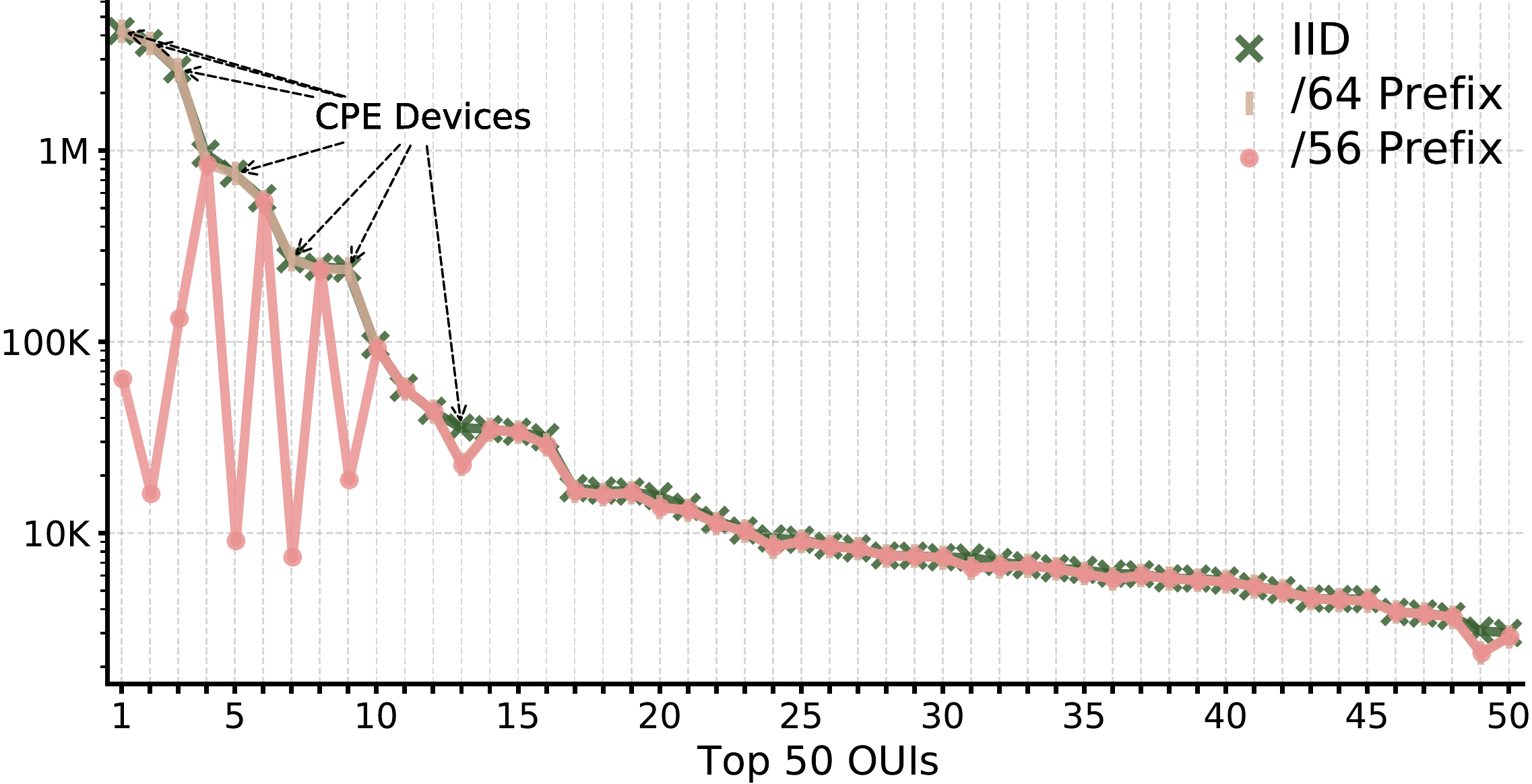}
	\caption{OUI popularity. Note that the y-axis is log-scaled.}
	\label{fig:isp-OUIs}
\end{figure}

As mentioned in Section~\ref{sec:datasets}, the ISP assigns /56
addresses to each subscriber line. In Figure~\ref{fig:isp-all-EUI64}
(right), we illustrate the number of prefixes that contain devices that
use EUI-64, non-EUI-64 devices that use privacy extensions, and the
prefixes that contain both types of addresses(i.e., dual-type
prefixes). In total, we observed at least one EUI-64 device in around
2.68 million /56 prefixes out of 11.3 million /56 prefixes. Thus, the
number of affected /56 prefixes accounts for about 22.2\%. Note that
the vast majority (more than 93\%) of the host prefixes with EUI-64
devices also host non-EUI-64 devices as well.  This shows that the
presence of privacy-violating EUI-64 addresses impacts a substantial
portion of ISP subscribers.   Even within a day, it
  is still possible for prefix rotation to happen for some subscriber
  lines. We can detect these rotations for EUI-64 using prefixes by
  tracking the IIDs across multiple /56 prefixes. We observed that
  only less than 13\% of the EUI-64 using /56 prefixes had prefix
  rotation within a day. Hence, if the same IID is observed across multiple /56
  prefixes, we count the prefixes only once. For non-EUI-64 prefixes, we
  cannot track them across prefixes after prefix rotation, which is
  precisely the purpose of using IPv6 privacy extensions.

\subsection{Popularity of EUI-64 Manufacturers}
\label{sub:quantify}

Based on the IPv6 address for devices that use EUI-64 addresses, we
analyze the device manufacturer using the OUI (see
Section~\ref{sec:methodology}). In total, we find devices with 1216
unique OUIs from 1113 distinct manufacturers.  In
Figure~\ref{fig:isp-OUIs}, we show manufacturers sorted by popularity
(\ie number of unique IIDs).  We focus on the top 50 manufacturers as
these are responsible for more than 99.1\% of all IIDs.  A closer
investigation shows that 6 out of the top 10 are CPE
manufacturers. The rest in the top 10 are IoT, smart TV, mobile
devices, home appliances, and data storage manufacturers.

\begin{figure} [t]
	\captionsetup{skip=.25em,font=small}
	\includegraphics[width=1\linewidth,valign=t]{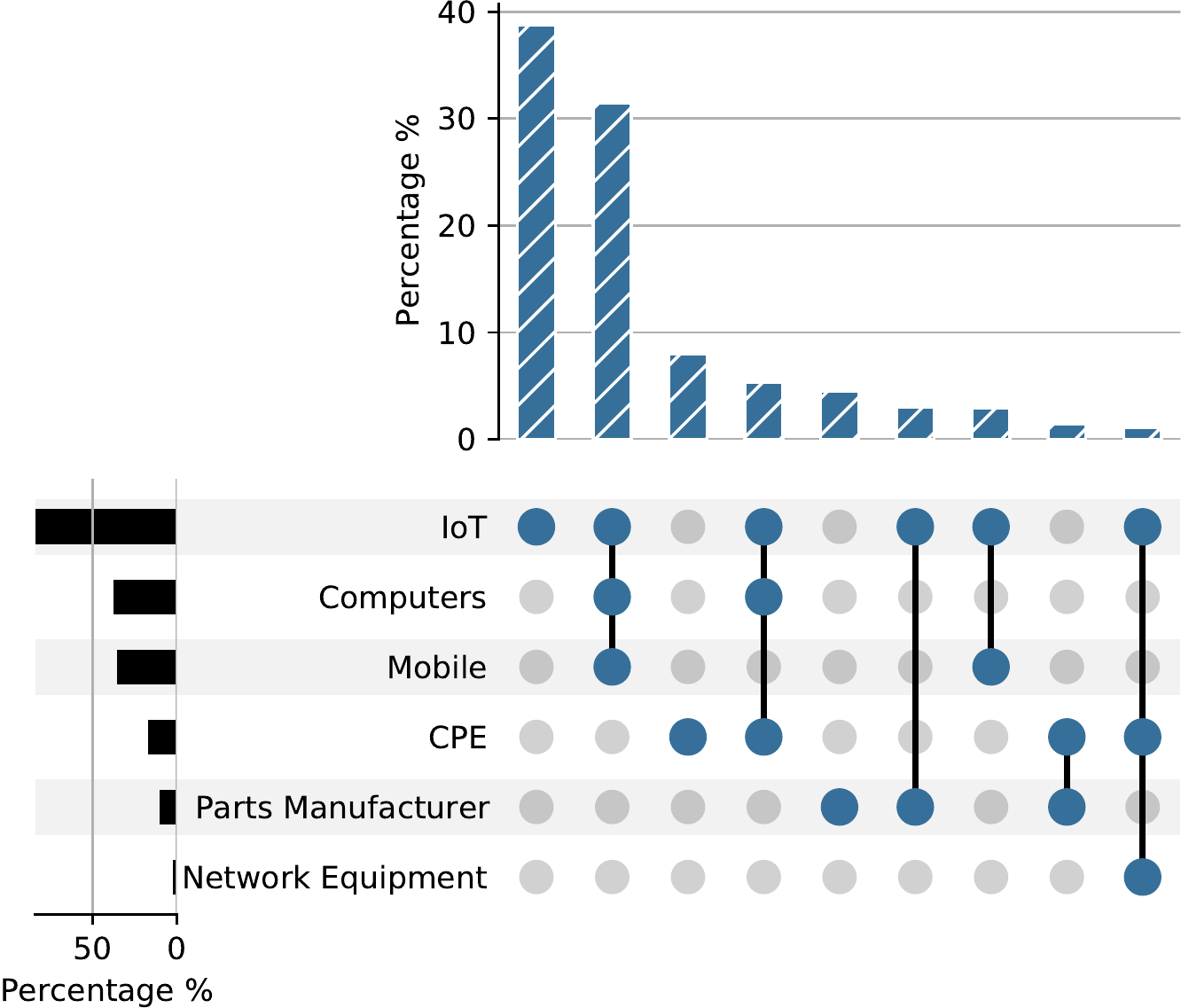}
	\caption{EUI-64 addresses mapped to different device categories. Percentage corresponds to number of /56 prefixes.}        
	\label{fig:isp-unique-subscribers}
        \vspace{-0.3cm}
\end{figure}

Interestingly, the number of covered /56 prefixes or even /64 ones is
almost identical with the number of IIDs in most cases. This means
that it is expected to be one device from each manufacturer in each
/56 or /64 in our dataset. A striking difference is the case of CPEs,
where the number of /56 prefixes is substantially lower than the /64
prefixes and the corresponding IIDs.  We attribute this to two
reasons.  First, the WAN (upstream-facing) interfaces of the CPEs in
the ISP typically do not share the same /56 prefix as the devices at
home, i.e., the periphery prefix is different from the end-user prefix
shown in \Cref{fig:iot-privacy-leakage}. We confirm this with multiple
users of the ISP.  Second, the IPv6 address of the WAN interface of
CPEs are concentrated within a relatively small number of prefixes
that it seems the ISP uses for exactly this purpose. Thus, in our methodology, the
  IPv6 address of the CPE is not sufficient to track the devices at
  home. On the other hand, it is possible to use this information to
  defeat the privacy of devices at home with active
  measurements~\cite{rye2021follow}.

Based on these insights, we re-estimate the number of affected
prefixes by differentiating between periphery subnets used by CPEs and
end-user subnets used by devices at home. We find that around 2.23M
prefixes out of a total of 2.6M EUI-64 prefixes are end-user prefixes.
Therefore, about 19\% of all 11.3M /56 prefixes are at risk of privacy
leakage.

\subsection{EUI-64 Manufacturer Categorization}

To understand what type of devices contribute the most to the leakage
of users' privacy due to EUI-64, we characterize the business model
and products of the associated manufacturers. Thus, we manually
visit the website of top 100 manufacturers found by our
method. We consider the following business types and any combinations:
IoTs, computers, mobile devices, CPEs, part manufacturers,
and network equipment manufacturers (see \Cref{sec:categories} for a
detailed description).  We assign a weight to each manufacturer with
the associated coverage of /56 prefixes. Then, we aggregate the
weights for the manufacturers of the same type.

\begin{figure} [!bpt]
	\captionsetup{skip=.25em,font=small}
	\includegraphics[width=0.9\linewidth,valign=t]{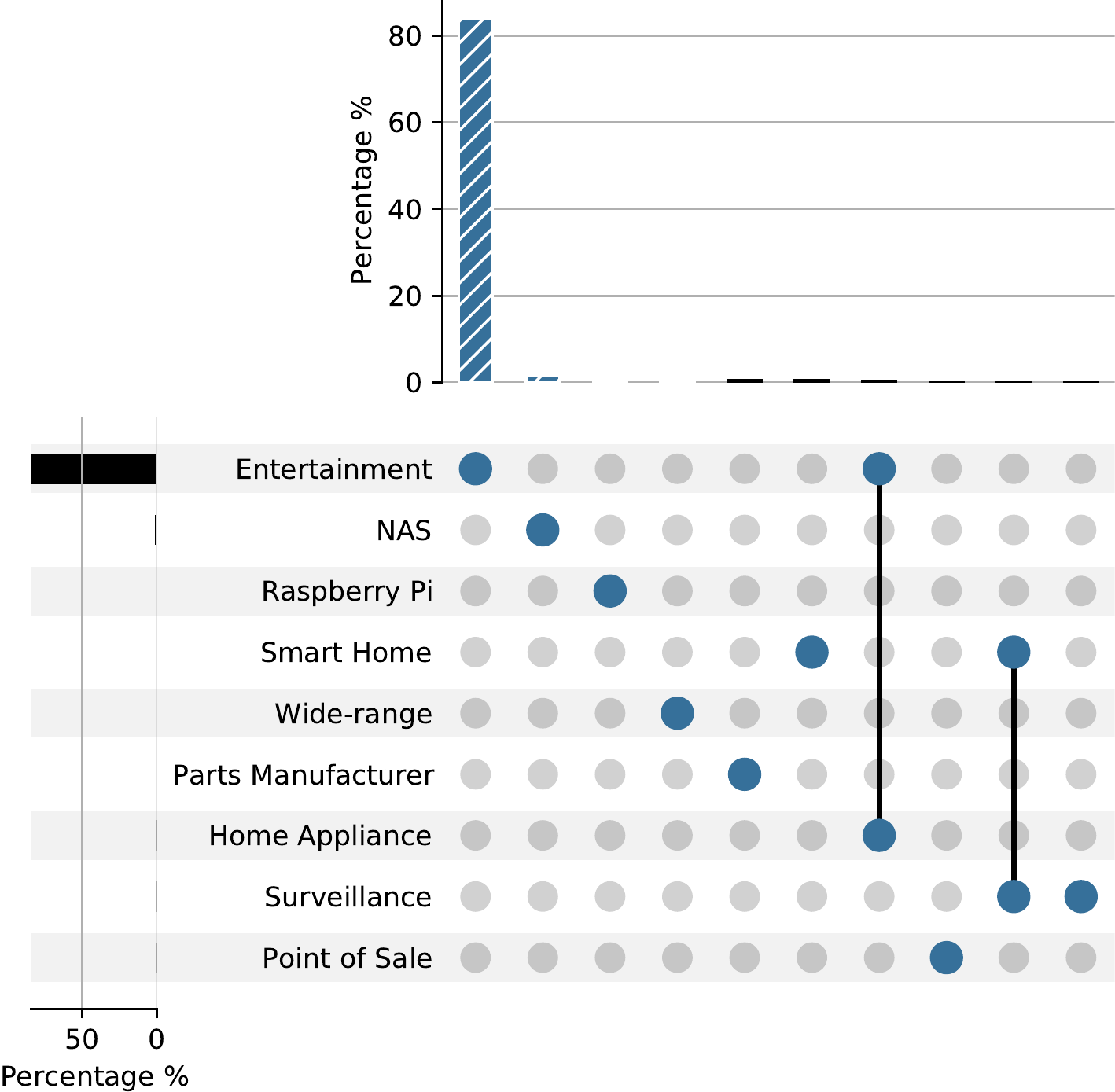}
	\caption{Composition of IoT-only manufacturers. Percentage corresponds to the number of /56 prefixes.}
	\label{fig:isp-iots}
        \vspace{-0.3cm}
\end{figure}

As shown in Figure~\ref{fig:isp-unique-subscribers}, around 39\% of
the prefixes that host EUI-64 devices, contain products from
manufacturers that only produce IoT devices. The second most popular
category, that accounts for 32\%, are devices by manufacturers active
in different product lines that include IoT devices, computers, mobile
devices. All other categories account for 8\% or less.  Thus, the
large majority of subscribers with EUI-64 devices are IoTs or likely
IoTs. To our surprise, a large number of subscribers host computers,
mobile phones and other equipment that also uses EUI-64. Although
large vendors, e.g., Apple, by default enable privacy extensions in
their products~\cite{Apple:blog:IPv6-security}, it seems that other
popular vendors do not. This could be related to some operating
systems not enabling privacy extensions by default.

\subsection{EUI-64 Use Among IoT Devices}

Next, we focus on the IoT devices that contribute the most to the
leakage of users' privacy. We take a conservative approach by only
considering manufacturers which exclusively produce IoT devices. We
manually investigate their product line and further categorize their
products as follows: entertainment (that includes smart TV, voice
assistants, streaming devices, media players), network attached
storage (NAS), Raspberry Pi, smart home equipment, IoT parts
manufacturer, home appliances, surveillance devices, point of sale
devices, and varied products (that include multiple categories). See
\Cref{sec:iot-categories} for a detailed description of these
categories.  As \Cref{fig:isp-iots} shows, the most popular
category is entertainment IoT devices, which cover more than 85\% of
all /56 prefixes with only IoT devices. In this category, we
identify more than 19 popular manufacturers. If this relatively small
number of manufacturers had adopted best common practices to enhance
IPv6 privacy, EUI-64 privacy leaks could have been substantially reduced.

\begin{figure} [!bpt]
	\captionsetup{skip=.25em,font=small}
	\includegraphics[width=.9\linewidth,valign=t]{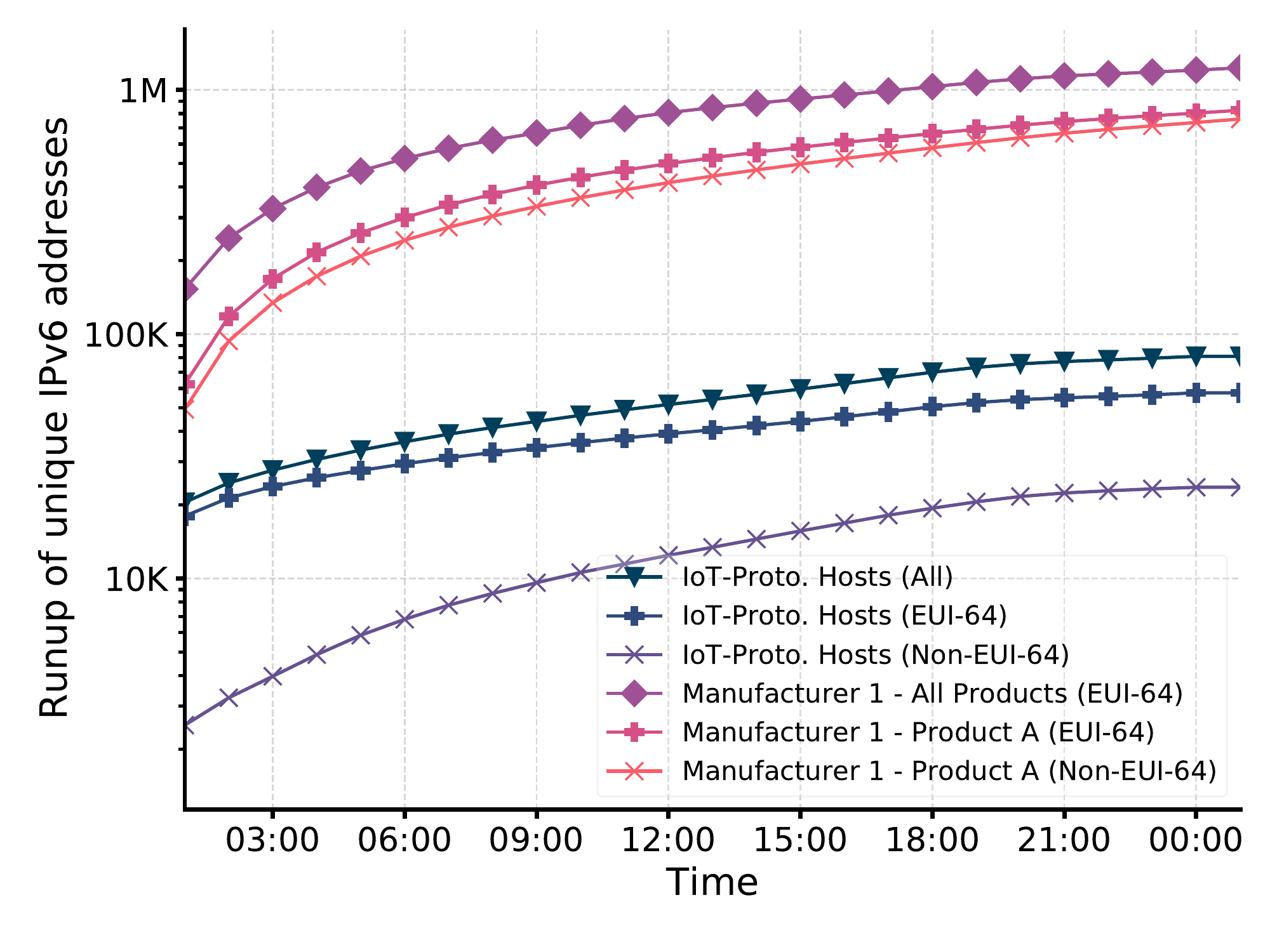}
\vspace{-1em}	
\caption{Prevalence of EUI-64 in IoT devices and a case study for a
popular IoT product. Note that the y-axis is log-scaled.}
	\label{fig:isp-amz-ips}
\vspace{-1em}
\end{figure}

However, even at the level of a manufacturer, it is possible that different products
or product versions have different behavior when it comes to
privacy leakage. To assess how common this is, we consider the top contributor
of EUI-64 IoT devices in our dataset (``manufacturer 1''). Using the methodology that we introduced
and validated in our previous work~\cite{IMC2020-IoT}, we annotate the
products of this IoT manufacturer based on the contacted destinations addresses.
We utilize the destination information to annotate the most popular
IoT product of this manufacturer (``product A''), and we also infer if
a specific device uses EUI-64 or not, based on the IPv6 address. In
Figure~\ref{fig:isp-amz-ips}, we show the cumulative unique number of
IPv6 addresses for all the products with EUI-64 of the manufacturer
and the number of devices with product A with and without EUI-64 with
hourly updates. A first observation is that, as expected, within 24
hours, the number of IPv6 addresses that host the EUI-64 devices as
well as the IIDs of this manufacturer converge to around 1.2 million
and 650k IPv6 addresses for the total and product A, respectively. The
number of IPv6 addresses and IIDs that do use and do not use EUI-64 is
similar for product A. Even though some of the devices belonging to
product A have adopted IPv6 privacy extension, either by updates or
because of newer models, the majority of these devices still have
the potential to leak user privacy.

Unfortunately, it is not easy to generate signatures for all IoTs based on the
visited destinations because the IoT devices have to be purchased, and
communication data has to be collected in a lab over longer periods of
time~\cite{IMC2020-IoT}. On the other hand, IoT-specific protocols
such as MQTT~\cite{MQTT:doc:MQTT-standard} are popular among many IoT
manufacturers. Indeed, we notice that port TCP/8883, i.e., the IANA-assigned port
for MQTT, is among the top 10 ports by our top manufacturers (see
Figure~\ref{fig:port-popularity} in Appendix~\ref{sec:port-popularity} for a
detailed view of ports used by different manufacturers). Hence, we use this
activity as a proxy to infer what is the percentage of IoT-devices that use
EUI-64 vs. any other MQTT activity that does not use EUI-64. In addition, we
confirm that more than 95\% of these devices contact servers that are
exclusively used for IoT cloud
services~\cite{Amazon:blog:IoT-core,Google:iot-core}. Therefore, these devices
are highly likely to be IoTs. Our analysis in Figure~\ref{fig:isp-amz-ips} shows
that, indeed, more than 83\% of the devices that communicate using the common IoT
protocol MQTT are also using EUI-64. This is another indicator of the rampant privacy-violating
practice of using EUI-64 addresses among IoT devices.

\begin{figure} [!bpt]
	\captionsetup{skip=.25em,font=small}
	\includegraphics[width=0.8\linewidth,valign=t]{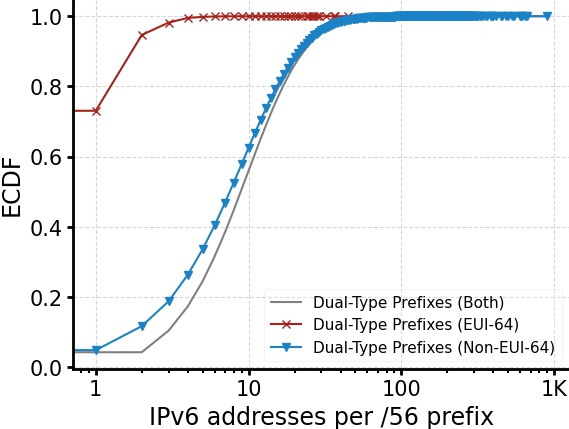}
    \caption{IPs in prefixes with both EUI-64 and non-EUI-64 IPs, \ie dual-type prefixes. Number of non-EUI-64, EUI-64, and both types of IPs in dual-type prefixes.
X-axis is log-scaled.}
    \label{fig:isp-eui64-vs-noneui64}
    \vspace{-0.5cm}
\end{figure}

\subsection{Collateral Privacy Leakage}

In this section, we turn our attention to the popularity of EUI-64 devices in
end-user prefixes. As shown in Figure~\ref{fig:isp-eui64-vs-noneui64}, we
typically only find one or two EUI-64 devices per end-user prefix.  Indeed, more
than 90\% of end-user prefixes that host both EUI-64 and non-EUI-64 devices,
i.e., are dual-type prefixes, have two or fewer EUI-64 devices. Only about 1\%
of dual-type prefixes host more than five EUI-64 devices. Recall, from
Figure~\ref{fig:isp-all-EUI64}, more than 93\% of all end-user prefixes with
EUI-64 devices also host non-EUI-64 devices. 

\begin{figure} [!bpt]
        \captionsetup{skip=.25em,font=small}
        \includegraphics[width=\linewidth,valign=t]{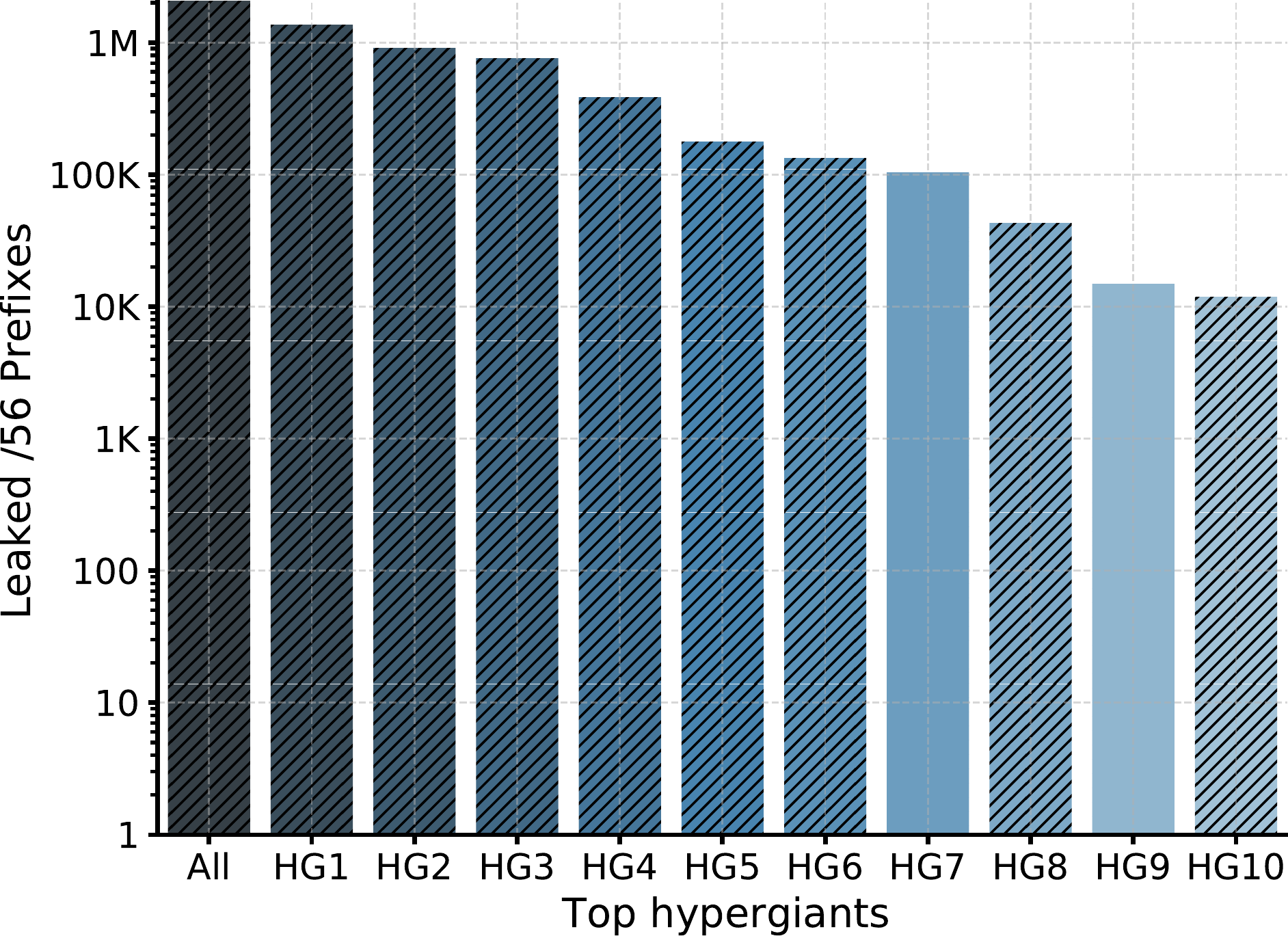}
		\caption{Per hypergiant number of /56 prefixes vulnerable to privacy. Each of these prefixes contains at least one EUI-64 address and one non-EUI-64 address.}
        \label{fig:isp-hypergiants}
        \vspace{-1em}
\end{figure}

Also, in Figure~\ref{fig:isp-eui64-vs-noneui64}, we see that number of non-EUI-64 addresses
in dual-type prefixes is larger than the number of EUI-64 addresses.
However, a single EUI-64 device is sufficient to leak user privacy to a third
party if both this device and a non-EUI-64 device contact the same destination.
To understand how probable this collateral privacy leakage is, first, we analyze
the popular applications that are contacted by EUI-64 devices. Our analysis shows
that these devices contact popular applications, e.g., Web (port 443,
80), DNS (port 53), NTP (port 123). For details about the popularity of ports
for the top EUI-64 manufacturers, we refer to Figure~\ref{fig:port-popularity} in
Appendix~\ref{sec:port-popularity}. This is alarming, as other devices that use
IPv6 privacy extensions also contact these ports. To estimate the collateral
damage, we count the number of dual-type prefixes where EUI-64 and non-EUI-64
devices contact the same third-party provider. \Cref{fig:isp-hypergiants} shows
the number of end-user dual-type prefixes which can be tracked over time by
common hypergiants \cite{bottger2017hypergiant}.  We find that in total, two
million end-user prefixes (around 17\% of the total end-user prefixes) 
are affected by this collateral privacy leakage, with
the top hypergiants, i.e., HG1, HG2, and HG3, being able to longitudinally
de-anonymize prefix rotation efforts by the ISP. Alarmingly, users do not even need to log in or visit the websites of these hypergiants to be tracked.
Tracking can simply happen by accessing one of their services, e.g., loading ads or static files.
Some of these hypergiants run popular public DNS services and online advertising platforms that make them
very attractive as a destination. 
A recent study also shows that services such as NTP can collect a vast number of
IPv6 addresses~\cite{bruns2020network}, thus, breaking IPv6 privacy when
sufficient conditions are in place, as we describe in our
methodology~(cf. \Cref{sec:methodology}). We note that this form of
tracking can not only be facilitated by hypergiants but also at
major aggregation points in the network, such as peering locations,
Internet exchange points, transit providers, and large data centers.

\section{Discussion} %

\noindent{\bf Vendor Self-regulation:} Hardware vendors should adequately test
their products and make every effort to protect the privacy of their consumers,
as currently, there is a gap in legislation regarding IPv6 privacy. This
includes all the involved parties, from chip manufacturers to product
integrators, software companies, and ISPs. For software companies, \eg operating
system distributors, it is important to enable IPv6 privacy extensions by
default. Unfortunately, at the time of writing, many Linux
distributions %
do not activate privacy extensions by default.
Products using Linux derivatives in their software are likely unknowingly
putting their users' privacy at risk. This could be related to the fact that
the original privacy extensions specification~\cite{rfc4941} contained a recommendation to
deactivate them by default. The current standard~\cite{rfc8981} does
not contain this recommendation anymore. We, therefore, recommend that all
IPv6-capable software stacks enable IPv6 privacy extensions by default.
We are in contact with hardware vendors to make them aware of this issue.

\noindent{\bf Privacy Badges:} The average user is not a privacy expert
when purchasing or operating smart home appliances or other Internet-connected
devices. Although the end-user may be aware of privacy risks when
using such devices, we can not expect end-users to perform experiments to validate which
devices use privacy extensions and which do not. The consumer unions and
regulators, \eg the FCC and FTC in the US and the European Commission in the EU could
require vendors to certify their products for IPv6 privacy compliance. These
badges could affirm the compliance of a product with the relevant future
legislation, similar to other certifications, \eg health, safety, and
environmental protection standards~\cite{CE-mark}.

\noindent{\bf The Role of the ISP:} ISPs should continuously improve the
privacy that they provide to their customers and could also inform them about
potentially privacy risky products in the market and their home network upon
customer request.
Another possibility would be to introduce a NAT in ISP IPv6 client networks.
This would, however, break the end-to-end principle---a primary design goal of IPv6 \cite{rfc4864}.
Therefore, we refrain from recommending NAT as a practical workaround.
Finally, ISPs should also check CPEs for privacy risks before shipping
them massively to their customers. 

\section{Conclusion}\label{sec:conclusion} %

In this paper, we show a new way to defeat IPv6 privacy even when the ISP does
prefix rotation. %
We find that a single device that uses EUI-64 can be leveraged as a tracking
identifier for devices in the same end-user prefix.  
Our analysis shows that up to a 19\% of end-user prefixes in a large ISP can
face IPv6 privacy leakage, and up to 17\% of them can be monitored by third
parties, primarily hypergiants. Closer investigation unveils that IoT devices
and popular manufacturers contribute the most to this type of IPv6 privacy
leakage. We propose that vendors should enable privacy extensions by default and
that regulatory intervention is necessary to protect users' privacy. In the
future, we continue to monitor prevalence of EUI-64 devices, and we extend our study by collaborating with other ISPs.

\section*{Acknowledgments}
We thank Marco Mellia and the anonymous reviewers for their valuable feedback.
This work was supported in part by the European Research Council (ERC)
Starting Grant ResolutioNet (ERC-StG-679158).

\bibliographystyle{plain}
\bibliography{paper}
\appendix

\vspace{3em}

\section{Appendix}

\vspace{1em}

\subsection{Analysis of Non-EUI-64 IPv6 Addresses}
\label{sec:hamming}

\begin{figure}[t]
	\includegraphics[width=1\linewidth,valign=t]{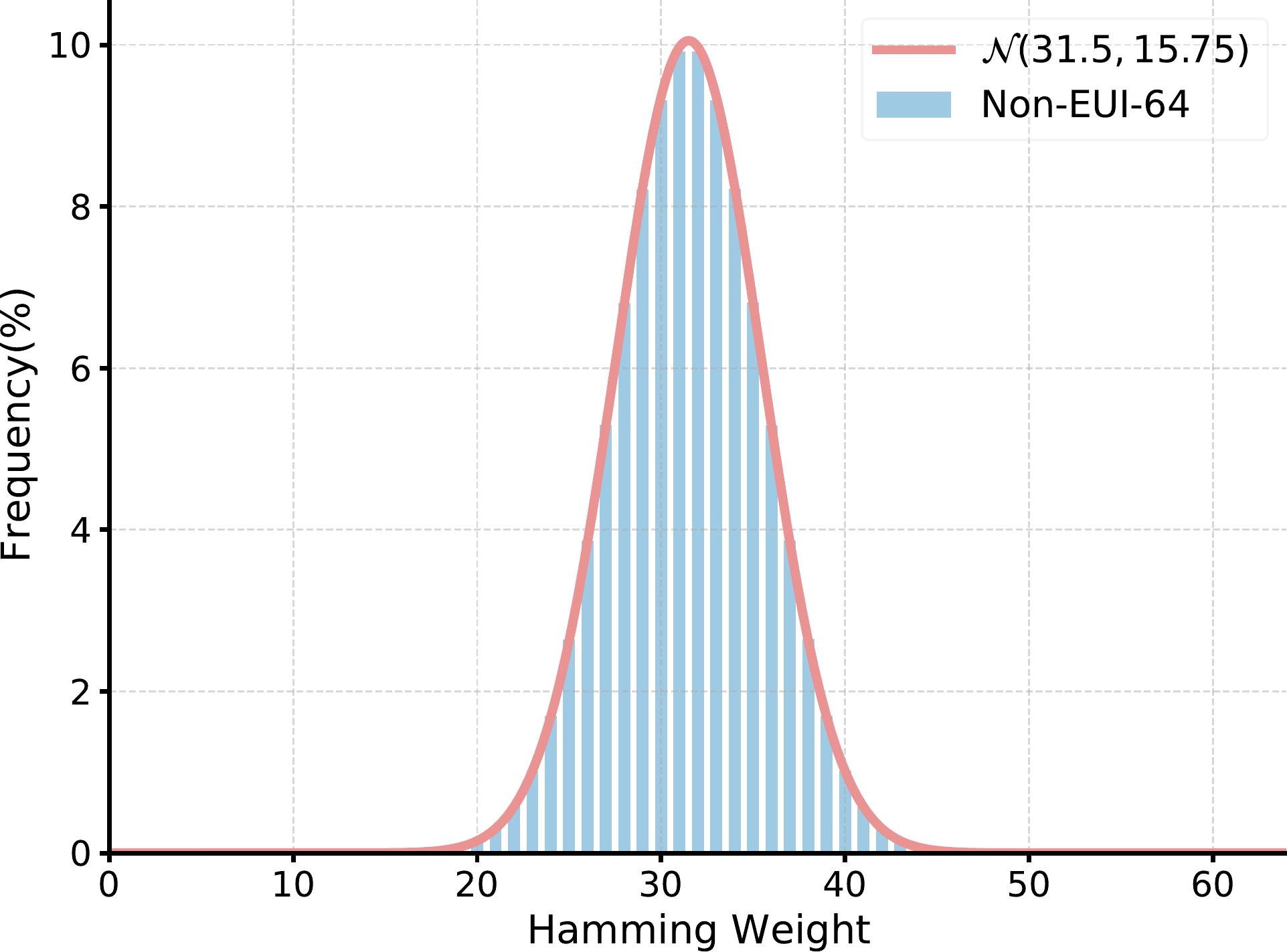}
	\caption{Hamming weight distribution of non-EUI-64 IIDs.}
	\label{fig:hamming-weight}
\vspace{1em}
\end{figure}

Non-EUI-64 addresses can be privacy extension addresses, addresses assigned via
DHCPv6, or also statically assigned addresses.  In order to understand how many
of the non-EUI-64 addresses are actually privacy extension addresses, we analyze
the interface identifier (IID) of all non-EUI-64 addresses.  We use the Hamming
weight, \ie the number of bits set to `1', to analyze the random nature of IIDs.
In completely random 64 bit IIDs, \ie the presence of privacy extensions, we
would expect exactly half of the bits being set to `1'.  Moreover, the central
limit theorem states that the sum of those independent IID Hamming weight
distributions tends toward a normal distribution.  In \Cref{fig:hamming-weight}
we show the Hamming weight distribution of these IIDs along with the normal
distribution shifted one bit to the left due to the universal/local bit.  As can
be seen, the non-EUI-64 Hamming weight distribution perfectly matches the normal
distribution.  Consequently, non-EUI-64 addresses in our dataset are in fact
privacy extension addresses.

\subsection{Device Categories}\label{sec:categories}

\begin{table}[!bpt]
\small
\begin{tabularx}{\columnwidth}{@{} l Y @{}} %
\toprule
Category & Description \\
\midrule
 IoT &  Manufacturers of internet-connected devices such as sensors, smart TVs, 
home appliances, security cameras, alarms, smart speakers, etc.  \\
 Computers  & Laptops, personal computers, and servers \\
 Mobile &  Mobile phones and tables \\
 CPE &   Devices supporting broadband technologies such DSL, cable modem, and 5G/4G hotspots. \\
 Parts Manufacturer &  Network interface cards, CPUs, memory modules, motherboards, WiFi modules, and chipsets that can be embedded into other devices. \\
 Network Equipment &  Routers, switches, access points, and firewalls. \\
 Gaming Console &  Internet connected devices primarily used for gaming.  \\
 Unknown &   Manufacturers that we were not able to find their website, or were not providing any information about the type of their products.  \\
 Virtual Machine & Vendors that develop virtual machine and hypervisor software.\\ 
 \bottomrule
\end{tabularx}
\caption{Description of device categories.}
\label{tab:device-cat}
\end{table}

\begin{figure*} [t]
	\includegraphics[width=.8\linewidth,valign=t]{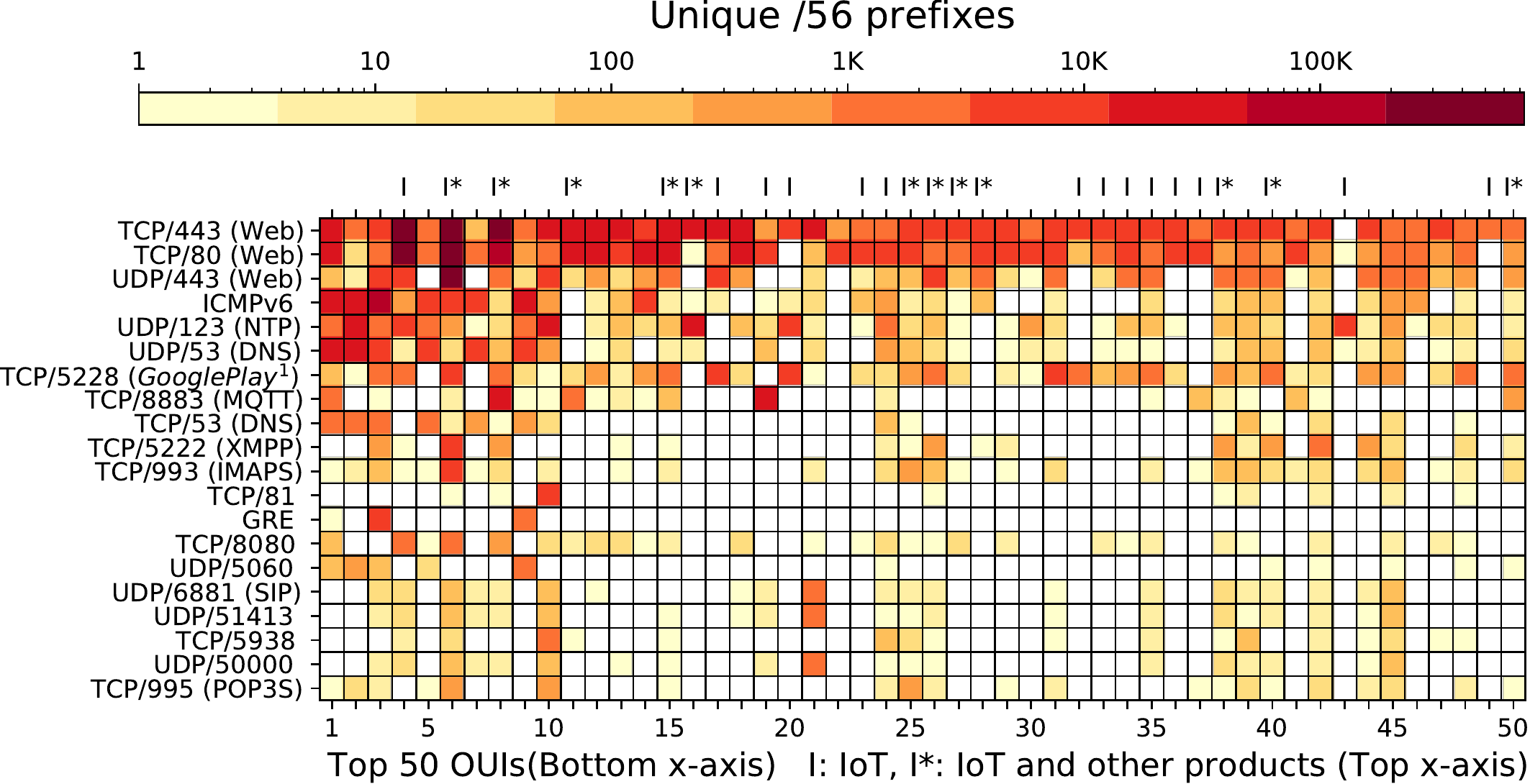}
	\caption{Heatmap showing used application ports for the top 50 OUIs. ¹The Google Play port is not officially registered with IANA.}
	\label{fig:port-popularity}
\end{figure*}

\begin{table}[!bpt]
\small
\begin{tabularx}{\columnwidth}{@{} l Y @{}} %
\toprule
Manufacturer Type & Description \\
\midrule
 Entertainment & Manufacturer of smart TVs, over the top streaming devices (OTTs), smart speakers, network-connected media players. \\
 Network Attached Storage & Internet-connected devices used for storing data. \\   
 Smart Home & devices such as smart plugs, light bulbs, door openers, alarms, and thermostats.  \\
 Varied & Manufacturers with a large portfolio of IoT devices, that not only includes all our categories but span beyond them. For example, robots, industrial devices, highly-specialized medical equipment, etc. \\   
 Parts Manufacturer & Chipsets, and modules tailored to be used specifically in IoT devices, e.g. 3G/4G, and Zigbee modules. Note, we tag a manufacturer in this category, only if it explicitly states that it produces IoT-specific modules and chipsets. \\
 Home Appliance & Washing machine, refrigerators, air conditioners, air purifiers, etc.   \\
 Surveillance & Security cameras and related surveillance equipment.  \\
 Point of Sale & Devices mostly used at retail stores for accepting payments. \\
\bottomrule
\end{tabularx}
\caption{Description of IoT manufacturer categories.}
\label{tab:manufacturer-cat}
\end{table}

By associating OUIs to their manufacturers, we can, for many OUIs, even identify
the type of device. The IEEE OUI database~\cite{mac-oui} contains details such
as the name and address of the company that registers an OUI. Depending on the
range and type of products of a manufacturer, it is possible to identify the
type of a device. For example, if we observe an OUI registered by a company
producing only wind turbines, the device generating the traffic is likely a wind
turbine. For this purpose, we visited the company's website that registered our
OUIs. We categorized their products into one or multiple categories. Some
companies produce generic products, e.g., network interface cards (NICs) installed in many
devices. In such cases, we mark their OUIs as Parts Manufacturers. Moreover, if
a company produces more than one product category, we assign their OUIs into all
those categories. \Cref{tab:device-cat} explains the different types of devices
categories that we used in our device classification.

\subsection{IoT Manufacturer Categories}\label{sec:iot-categories}

\Cref{tab:manufacturer-cat} explains the different types of IoT manufacturer categories that we used in our EUI-64 classification.

\subsection{Traffic Profile by Manufacturer}\label{sec:port-popularity}

Figure~\ref{fig:port-popularity} shows the popularity among the top 20 protocols utilized by
top 50 manufacturer devices that use EUI-64. These devices utilize protocols
that are also popular for other devices, like laptops, smartphones, etc. that
may use privacy extension. Thus, it is possible that devices using EUI-64 and
other that do not use EUI-64 contact the same CDNs (Web on ports 80
and 443), applications (Google play updates on port 5228 ~\cite{Google-play}, MQTT on port 8883), or other services (NTP on port 123, DNS on port 53).

\end{document}